\documentclass[runningheads]{llncs}
\usepackage[T1]{fontenc}
\usepackage{amsmath,amssymb,amsfonts}
\usepackage{algorithmic}
\usepackage{graphicx}
\usepackage{textcomp}
\usepackage{xcolor}
\usepackage{enumitem}
\usepackage{soul}
\usepackage{amssymb} 

\newif\ifreview
\reviewfalse 

\ifreview
\newcommand{\review}[2]{{\hl{[#1]} \color{red}#2}}
\else
  \newcommand{\review}[2]{#2}
\fi

\usepackage[capitalize]{cleveref}
\crefname{equation}{Eq.}{Eqs.}

\usepackage{tikz}
\usepackage{url}

\newcommand{\springernotice}{%
  \footnotesize
  \textcopyright\ 2027 The Author(s), under exclusive license to
  Springer Nature Switzerland AG.\\
  This is a preprint version. The final authenticated version is
  available online at:
  \url{https://doi.org/10.1007/978-3-032-36590-3_6}
}

\AddToHook{shipout/firstpage}{%
  \begin{tikzpicture}[remember picture,overlay]
    \node[
      anchor=south west,
      xshift=1.0cm,
      yshift=0.7cm
    ] at (current page.south west) {%
      \parbox{0.9\paperwidth}{\raggedright\springernotice}%
    };
  \end{tikzpicture}%
}

\begin{document}

\title{Model-Based Retargeting to Many-Core CPS: Simulink-to-OpenCL Workflow}
\titlerunning{Model-Based Retargeting to Many-Core CPS}
\author{Ryuga Chiba\inst{1} \and
Hiroshi Fujimoto\inst{2} \and
Takuya Azumi\inst{1}}
\authorrunning{R. Chiba et al.}
\institute{Graduate School of Science and Engineering, Saitama University \and
eSOL Company Ltd.}

\maketitle
\begin{abstract}
This paper addresses the software portability gap between Model-Based Development (MBD) and advanced many-core execution for Cyber-Physical Systems (CPS). We present a workflow-preserving retargeting approach for Simulink-based CPS applications with candidate-wise data parallelism to OpenCL-based many-core processors. Rather than manually rewriting models for new platforms, our toolchain uses MathWorks GPU Coder to extract data-parallel CUDA code, which is then translated into OpenCL host and device code via a custom framework. The conversion handles syntax rewriting, API emulation, and platform-specific argument packing. We deployed this workflow for a computationally intensive Frenet-frame trajectory planner on the Kalray MPPA Coolidge2. The results demonstrate the feasibility of a workflow-preserving retargeting pipeline for the evaluated CPS workload and platform.

\end{abstract}

\section{Introduction}\label{sec:intro}
The software engineering of Cyber-Physical Systems (CPS)~\cite{alur2015principles,derler2012cyber}, such as autonomous driving~\cite{8443742}, increasingly relies on Model-Based Development (MBD)~\cite{broy2007,miura2021cosam} to facilitate early algorithm verification and ensure safety. However, as the underlying computational platforms diversify toward many-core architectures, software portability and parallel performance have emerged as critical challenges. In many cases, commercial code generators tightly lock models into specific vendor architectures (e.g., NVIDIA CUDA). This reveals a gap between model-level abstraction and platform-level execution in CPS, particularly when moving beyond GPU-centric toolchains.

To bridge this gap, this paper proposes a model-based retargeting workflow thatsupports the deployment of Simulink models on OpenCL-based many-core platforms for CPS applications~\cite{burgio2017software}. Our methodology leverages GPU Coder to extract data-parallel CUDA code, which provides a structured host/device representation suitable for systematic retargeting. Using a custom translation toolchain, this intermediate code is subsequently converted into OpenCL-compatible code without requiring computationally heavy abstract syntax tree (AST) analysis. 
This approach supports the retargeting of Simulink-generated CPS workloads
to OpenCL-based many-core platforms such as the Kalray MPPA Coolidge2~\cite{nozaki2025dedicated,yabe}, without disrupting the established MBD simulation and verification workflow.

To investigate this problem, we use a Frenet-frame trajectory planner~\cite{werling2012optimal,werling2010} as a representative candidate-parallel CPS workload. This choice is not intended to restrict the scope to a specific application, but to provide a concrete instance for analyzing model structure, translation mechanisms, and runtime behavior.

Through empirical evaluation on the Kalray MPPA Coolidge2 platform, we analyze how the generated code meets an assumed planning-cycle budget and what factors determine its end-to-end execution behavior. We hypothesize that retargetability is determined by three factors:
\begin{enumerate}
    \item model structure that exposes data-parallel computation,
    \item preservation of generated code structure across platforms, and
    \item system-level execution behavior including runtime overheads.
\end{enumerate}

The primary contributions of this paper are as follows:
\begin{itemize} 
 \item We present a workflow-preserving retargeting approach for Simulink-based CPS applications whose computationally intensive logic is expressed in MATLAB Function blocks and whose dominant workload can be decomposed into independent candidate-wise or data-wise computations.
 \item We demonstrate an automated data-parallel mapping at the trajectory-candidate level for many-core architectures.
 \item We conduct an end-to-end evaluation of the proposed workflow on a physical many-core platform, confirming that the assumed planning-cycle budget (100 ms) is met for a workload of 4,000 trajectory candidates while examining scalability.
\end{itemize}

Based on this perspective, this study addresses the following research questions: 
\begin{itemize}
\item [\textbf{RQ1.}] What model-level and generated-code structural conditions make a Simulink-based CPS workflow retargetable to an OpenCL-based cluster many-core processor?
\item[\textbf{RQ2.}] Which platform-aware translation mechanisms are necessary to preserve the generated host/device execution structure beyond a direct CUDA-to-OpenCL syntax rewrite?
\item[\textbf{RQ3.}] How do search-space size, planning-cycle budget, and end-to-end runtime overhead determine the timing behavior of model-generated candidate-parallel CPS workloads?
\item[\textbf{RQ4.}] How does end-to-end latency scale with the number of compute clusters, and which runtime/system factors bound this scalability?

\end{itemize}

The remainder of this paper is organized as follows.
Section~\ref{sec:systemmodel} introduces the system model and the associated assumptions.
Section~\ref{sec:design} details the proposed approach.
Section~\ref{sec:evaluation} reports the experimental results.
Section~\ref{sec:related} reviews related work.
Finally, Section~\ref{sec:conc} concludes the paper.

\section{System Model}\label{sec:systemmodel}
\begin{figure}[t]
  \centering
  \includegraphics[width=0.85\textwidth,keepaspectratio]{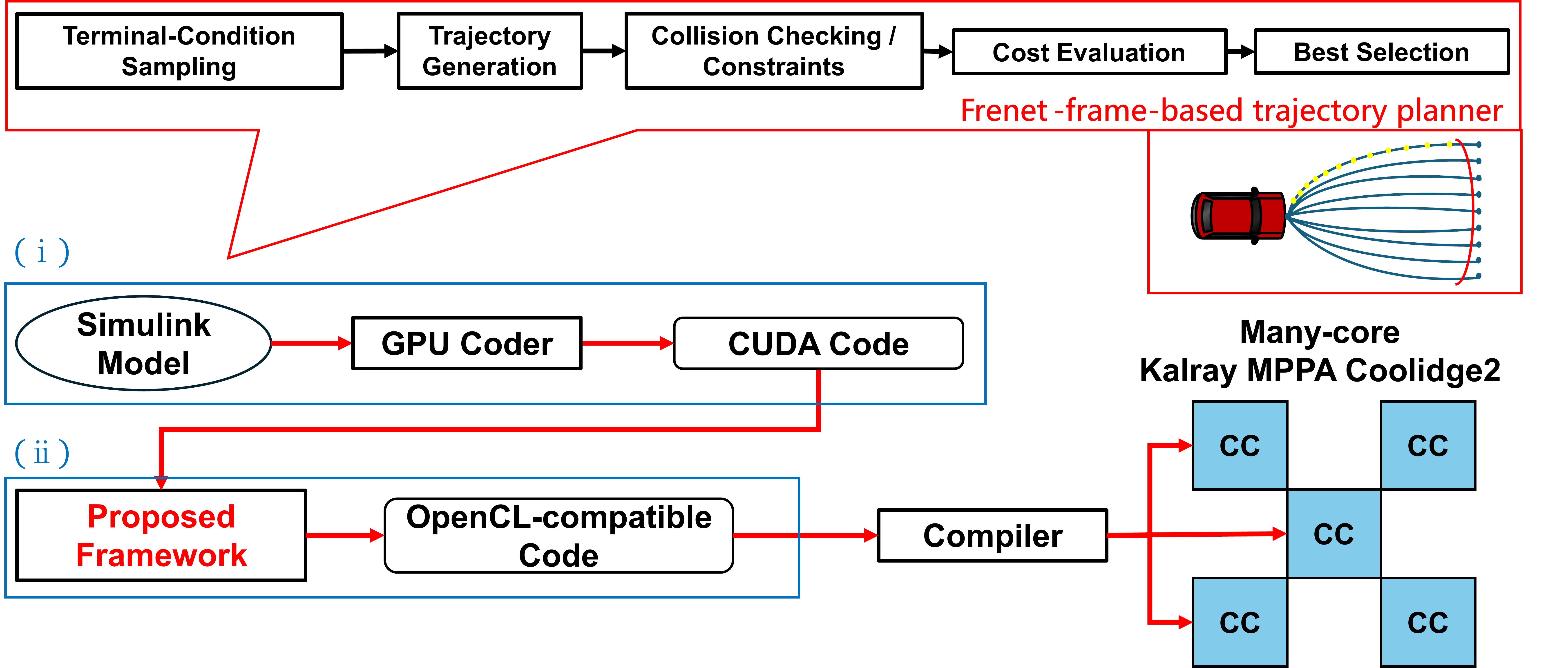}
  \vspace{-4pt}
  \caption{System model.}
  \label{fig:proposed}
\end{figure}
\vspace{-5pt}
\begin{figure}[t]
\centering

\begin{minipage}{0.48\columnwidth}
    \centering
    \includegraphics[width=\linewidth]{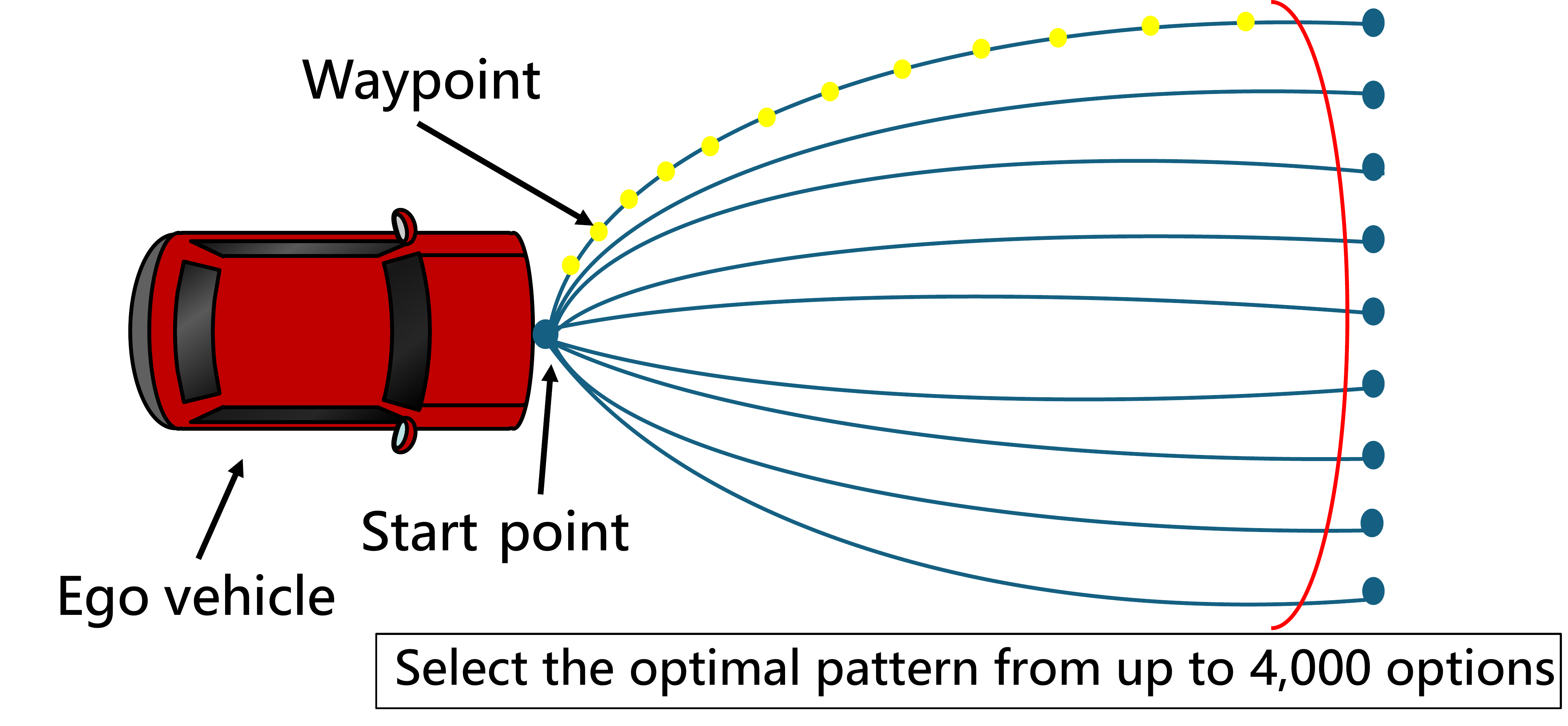}
    \vspace{-3pt}
    \caption{Frenet-frame-based trajectory planner}
    \label{fig:Frenet}

\end{minipage}
\hfill
\begin{minipage}{0.48\columnwidth}
    \centering
    \includegraphics[width=\linewidth]{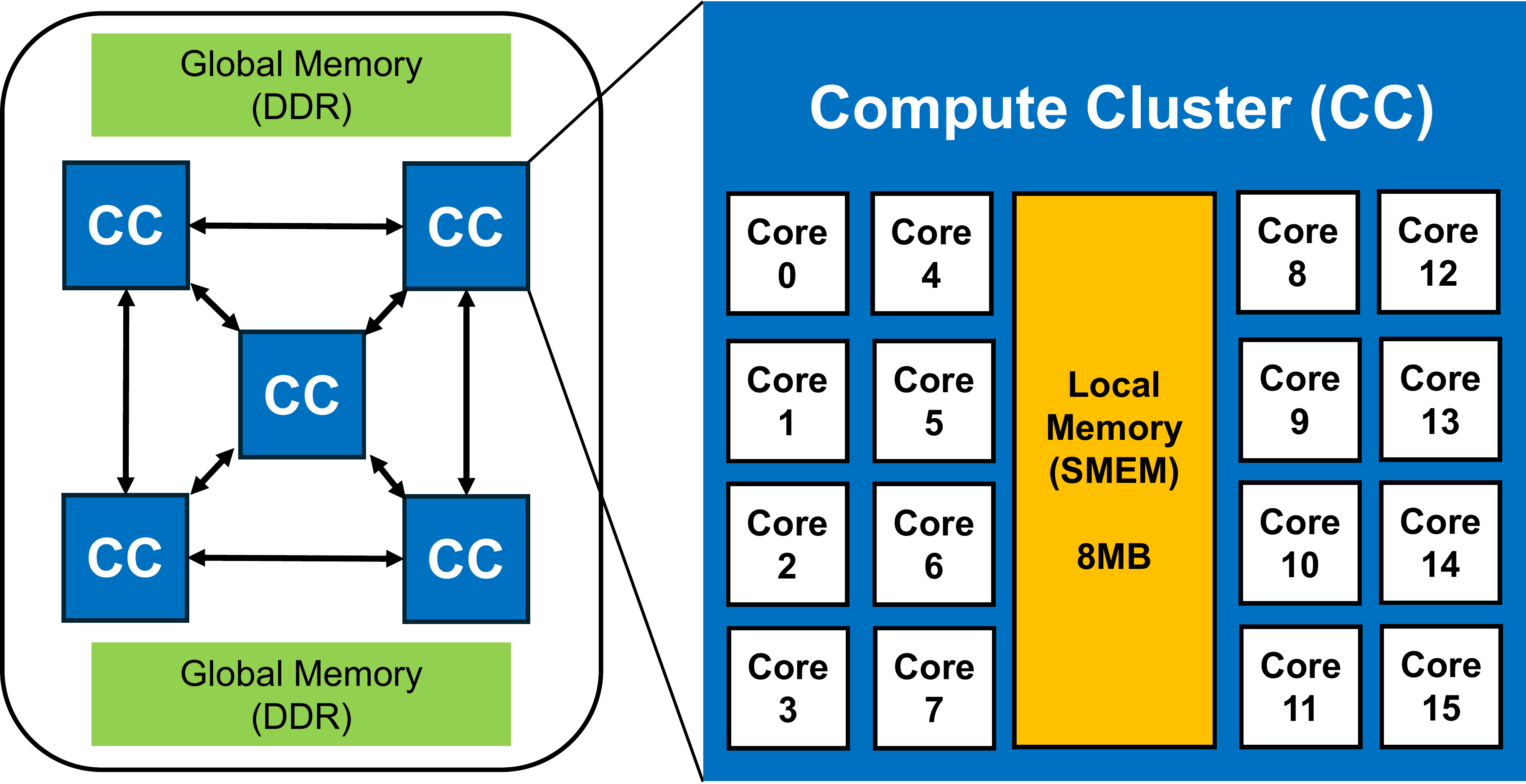}
    \vspace{-3pt}
    \caption{Architecture of Kalray MPPA Coolidge2}
    \label{fig:Coolidge}

\end{minipage}
\vspace{-14pt}
\end{figure}

An overview of the system model is shown in Fig. \ref{fig:proposed}.
This paper proposes a model-based parallelization and code-generation framework for deploying the Frenet-frame-based trajectory planner on OpenCL-based many-core processors.

The proposed framework consists of the following stages:
\begin{itemize}
  \item[(i)] CUDA code generation from a Simulink model using GPU Coder.
  \item[(ii)] CUDA-to-OpenCL conversion that produces OpenCL-compatible host code and kernel code.
\end{itemize}

\vspace{-4pt}
\subsection{Frenet-frame-based trajectory planner}

The Frenet-frame-based trajectory planner~\cite{werling2012optimal,werling2010} generates a large number of candidate trajectories by sampling terminal conditions in the Frenet coordinate system and selects the lowest-cost feasible trajectory as illustrated in Fig.~\ref{fig:Frenet}. The computation is independent across candidates, enabling data-parallel execution. In this study, $4{,}000$ candidates are generated from $40$ lateral offsets, $10$ time horizons, and $10$ target speeds. We adopt the standard formulation proposed by Werling et al.~\cite{werling2012optimal,werling2010} and refer the reader to that work for mathematical details.

\subsection{Simulink}\label{ssub:simulink}

Simulink is used to model the Frenet planner and serves as the starting point of the proposed workflow. To enable GPU Coder-based kernel generation, computationally intensive parts are implemented in MATLAB Function blocks. The resulting model allows candidate-wise computations to be expressed independently, forming the basis for data-parallel execution and subsequent code generation.

\subsection{GPU Coder}\label{ssub:GUPcoder}

GPU Coder converts MATLAB-based implementations into CUDA code for parallel execution. In Simulink models, CUDA kernels are primarily generated from MATLAB Function blocks, while other blocks are translated into host-side code. Therefore, computationally intensive parts must be implemented in MATLAB Function blocks to enable kernel generation.

\subsection{Many-core processors}\label{ssub:manycore}
The target execution platform of this study is Kalray MPPA Coolidge2 (Fig.~\ref{fig:Coolidge}), a cluster-based many-core processor. The processor integrates multiple compute clusters (CCs), each containing multiple processing cores and dedicated local memory connected via a network-on-chip. One CC can serve as the host running the OpenCL runtime, while the remaining CCs execute device kernels. Because each CC has its own local memory, execution performance depends on how workloads are partitioned across clusters and how inter-cluster communication overhead is managed.

\subsection{OpenCL}\label{ssub:opencl}
OpenCL is a vendor-independent parallel programming framework targeting heterogeneous devices. Its host–device execution model is similar to CUDA, enabling systematic translation while preserving data-parallel semantics. In this study, the Kalray MPPA Coolidge2 platform provides an OpenCL runtime for executing the converted code.

\review{R.1.1}{\subsection{Target Application Class and Assumptions}
The proposed workflow is not intended for arbitrary Simulink models. Rather, it targets a class of model-based applications that satisfy the following conditions. First, the computationally intensive part of the application must be implemented in MATLAB Function blocks so that GPU Coder can generate CUDA kernels from the model. Second, the dominant computation should be decomposable into many independent candidate-wise or data-wise tasks, which enables a one-work-item-per-task mapping in OpenCL. Third, the generated CUDA code must remain within the supported
GPU-Coder-generated subset handled by the proposed translator. Fourth, the kernel interface must be compatible with the target platform constraints, such as the kernel-argument limit on Kalray MPPA, which is addressed in this work by scalar-argument packing. 

Under these assumptions, the workflow can be applied not only to the Frenet trajectory planner studied in this paper, but also to other model-based applications that exhibit similar structural properties. Typical examples include candidate-based trajectory evaluation, independent scenario evaluation, and data-parallel state or cost updates. In contrast, models whose dominant computation is expressed mainly with basic Simulink blocks, or whose execution is dominated by strong sequential dependencies, are outside the intended scope of this workflow. 

The Frenet planner is used in this paper as a representative case study because it naturally satisfies these assumptions: trajectory generation, collision checking, and cost evaluation are independent across candidates, and the computationally intensive logic can be placed in MATLAB Function blocks for GPU-Coder-based kernel generation.
}
\vspace{-8pt}

\section{Design and Implementation}\label{sec:design}
\begin{figure}[t]
  \centering
  \includegraphics[width=\textwidth,keepaspectratio]{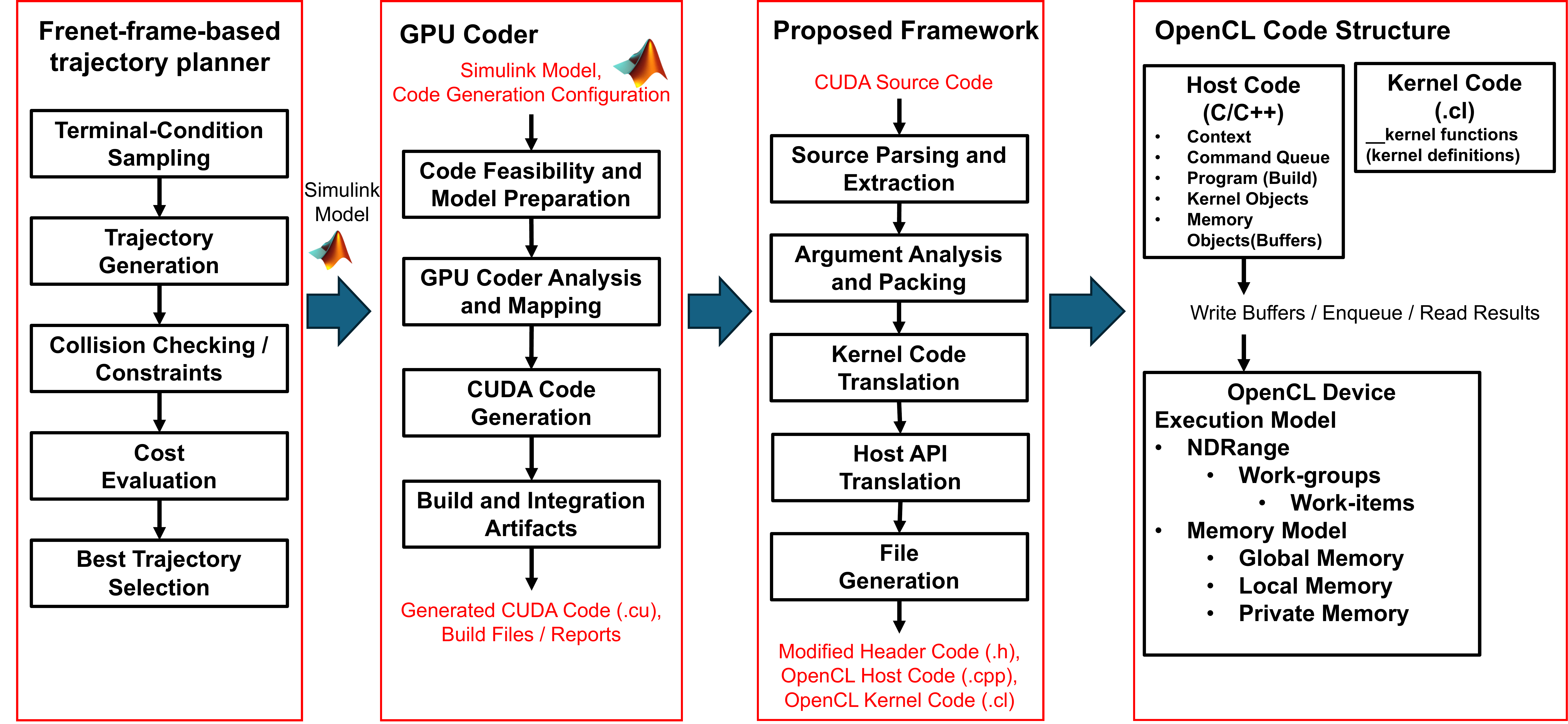}
  \caption{Workflow of the proposed CUDA-to-OpenCL conversion framework.}
  \label{fig:detail1}
      \vspace{-10pt}

\end{figure}

\begin{figure}[t]
  \centering
\includegraphics[height=0.20\textheight, keepaspectratio]{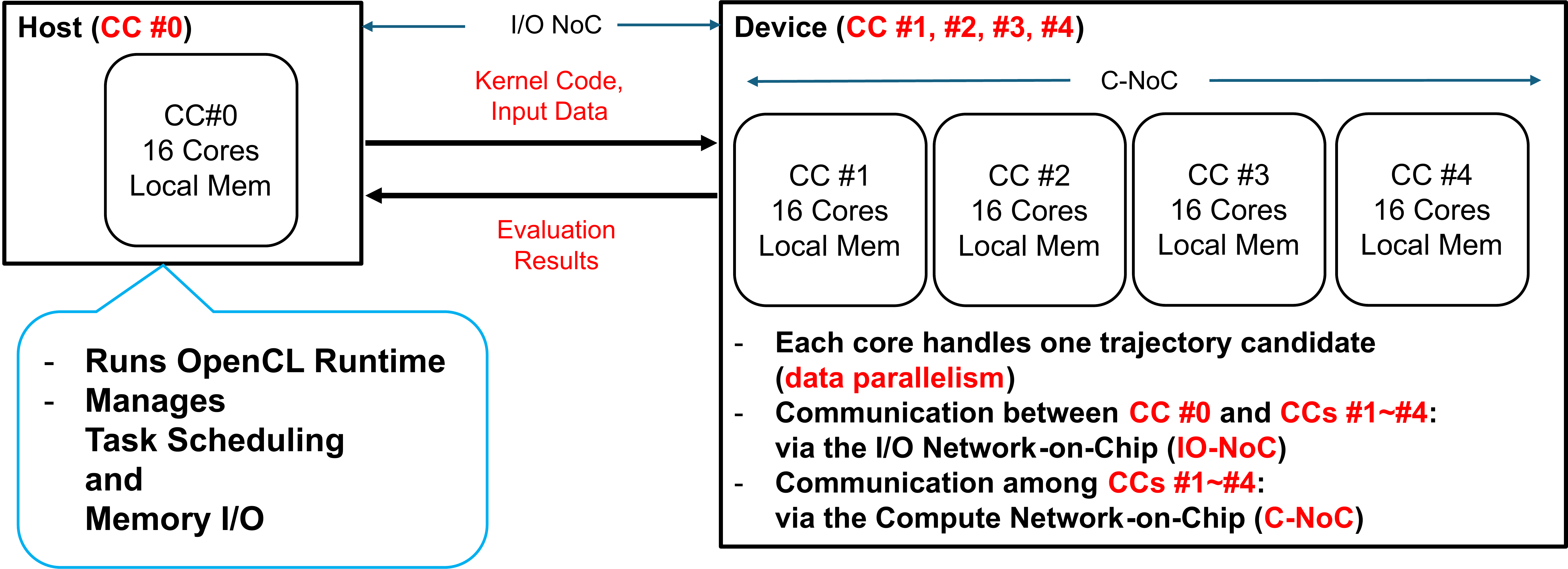}
    \caption{Cluster-level and work-item-level execution mapping on Kalray MPPA Coolidge2 platform.}  \label{fig:DH}
    \vspace{-10pt}

\end{figure}

This section describes the design and implementation of the proposed code generation and execution framework for deploying the Frenet trajectory planner on Kalray MPPA Coolidge2 platform. The framework converts CUDA code generated from Simulink models into OpenCL-compatible code and enables execution on OpenCL-based many-core architectures.

\review{R2.2}{
Rather than attempting to automatically generate executable OpenCL programs as conventional source-to-source translators do, the proposed framework focuses on rewriting CUDA-specific constructs into forms that are valid in standard C++ and OpenCL environments. The conversion ensures that the generated code can be compiled using the Kalray toolchain without manual modification. By converting CUDA code into OpenCL-compatible C++ code, the framework removes dependence on NVIDIA-specific toolchains and enables execution on heterogeneous many-core platforms such as Kalray MPPA Coolidge2 platform.
}
\vspace{-3pt}
\subsection{Rationale for Using CUDA as an Intermediate Representation}

CUDA is used as an intermediate representation rather than a final execution target. 
This choice is motivated by two reasons.

First, GPU Coder provides a practical and well-supported path from Simulink\slash MATLAB Function blocks to parallel code, producing CUDA programs with a clear host–device structure. This structure closely aligns with the OpenCL execution model, enabling systematic retargeting without modifying the original Simulink model.

Second, a direct Simulink-to-OpenCL code generation path that satisfies platform-specific constraints, such as those of the Kalray MPPA, was not available in our development environment. Existing approaches do not directly address platform-specific runtime integration and constraints, such as kernel argument limits and cluster-level execution mapping.

By using GPU-Coder-generated CUDA as an intermediate representation, platform-specific adaptations can be handled during translation while preserving the original model-based workflow.

\vspace{-3pt}
\subsection{Workflow Overview}

The proposed workflow starts from a Simulink model of the Frenet planner. CUDA code is generated using MathWorks GPU Coder, and the generated CUDA source is translated into OpenCL-compatible host and kernel code.
The overall workflow of the proposed framework is illustrated in Fig.~\ref{fig:detail1}.

The translation is implemented as a Python-based transformation tool. The converter rewrites CUDA-specific constructs into equivalent OpenCL-compatible forms while preserving the original program structure.

The conversion produces two types of output files:

\begin{itemize}
\item \textbf{OpenCL host code} (\texttt{.cpp}) responsible for OpenCL context creation, memory management, and kernel invocation.
\item \textbf{OpenCL kernel code} (\texttt{.cl}) containing translated device-side parallel computation.
\end{itemize}

The generated files are produced in a ready-to-build form and can be compiled directly using the Kalray MPPA toolchain.

The translated code is compiled and executed as follows.

\begin{description}[style=nextline,leftmargin=0pt]

\item[\textbf{Host Code Compilation}]
The C++ host code is compiled using \texttt{kvx-linux-g++}, a GCC-based cross-compiler targeting the KVX architecture.

\item[\textbf{Kernel Compilation}]
The OpenCL kernel code is compiled using \texttt{kvx-poclcc}, the POCL kernel compiler, producing precompiled \texttt{.pocl} binaries.

\item[\textbf{Execution}]
The compiled binaries are executed on Kalray MPPA Coolidge2 platform using the POCL OpenCL runtime.

\end{description}

\subsection{CUDA-to-OpenCL Translation}
\subsubsection{Syntax Rewriting}
The proposed converter translates CUDA code generated by GPU Coder
into OpenCL-compatible host and kernel code. The translation focuses
on rewriting CUDA-specific constructs into equivalent OpenCL
representations.

Table~\ref{tab:translation_rules} summarizes the main translation
rules applied by the converter.

\begin{table}[t]
\caption{CUDA-to-OpenCL Translation Rules}
\label{tab:translation_rules}
\centering
\small
\setlength{\tabcolsep}{4pt}
\renewcommand{\arraystretch}{1.1}

\begin{tabular}{|l|l|l|}
\hline
\textbf{Category} & \textbf{CUDA} & \textbf{OpenCL} \\
\hline

Kernel qualifier & \verb|__global__| & \verb|__kernel| \\
\hline
Thread index* & \verb|threadIdx.x| & \verb|get_local_id(0)| \\
\hline
Block index* & \verb|blockIdx.x| & \verb|get_group_id(0)| \\
\hline
Block size* & \verb|blockDim.x| & \verb|get_local_size(0)| \\
\hline
Grid size* & \verb|gridDim.x| & \verb|get_num_groups(0)| \\
\hline
Thread synchronization & \verb|__syncthreads()| & 
\verb|barrier(...)| \\
\hline
Kernel launch & \verb|<<<grid,block>>>| & 
\verb|clEnqueueNDRangeKernel| \\
\hline
Kernel argument binding & implicit & 
\verb|clSetKernelArg| \\
\hline
Pointer argument & \verb|Type *ptr| & 
\verb|__global Type *ptr| \\
\hline
Memory allocation & \verb|cudaMalloc| & 
\verb|clCreateBuffer| \\
\hline
Memory deallocation & \verb|cudaFree| & 
\verb|clReleaseMemObject| \\
\hline
Memory copy & \verb|cudaMemcpy| &
\begin{tabular}[c]{@{}l@{}}
\verb|clEnqueueWriteBuffer| \\
\verb|clEnqueueReadBuffer|
\end{tabular} \\
\hline
CUDA runtime header & \verb|cuda_runtime.h| & 
\verb|opencl_helper.h| \\
\hline
Memory qualifier & \verb|__shared__| & 
\verb|__local| \\
\hline
Function qualifier & \verb|__device__| & 
(Removed) \\

\hline
\end{tabular}

\vspace{2pt}
{\scriptsize $^{*}$The \texttt{.y} and \texttt{.z} dimensions are similarly translated to \texttt{(1)} and \texttt{(2)}.}
\vspace{-10pt}
\end{table}

The CUDA grid--block hierarchy is mapped to the OpenCL NDRange model: the global work size equals the product of grid and block dimensions, and the local work size equals the CUDA block size. This direct mapping preserves the data-parallel execution model generated by GPU Coder.

\subsubsection{Header and Runtime Emulation}

CUDA runtime APIs and data types are not available in standard C++ compilers. To compile GPU-Coder-generated host code without manual modification, the framework provides a header-based emulation layer (\texttt{opencl\_helper.h}). This header supplies compatible definitions for CUDA-specific types (\texttt{dim3}, \texttt{cudaError\_t}) and wraps CUDA memory APIs (\texttt{cudaMalloc}, \texttt{cudaMemcpy}) as thin functions that internally invoke the corresponding OpenCL operations. A pointer-to-\texttt{cl\_mem} mapping table is maintained so that the host code can continue to use raw pointers while the emulation layer manages OpenCL buffer objects transparently.
\subsubsection{Argument Packing for Hardware Constraints}

Deployment on the Kalray MPPA platform introduces a constraint on the maximum number of kernel arguments. Kernels generated from complex Simulink models often exceed this limit due to numerous scalar parameters.

To address this, the framework packs scalar parameters into a single structure, which is passed as one kernel argument. Array pointers are excluded from this packing to avoid invalid memory access when handling device memory (\texttt{\_\_global} pointers).

The translator generates consistent structure definitions for both host and device code. On the host side, parameters are assigned to the structure and transferred as a buffer, while the kernel unpacks the structure into local variables for computation.

This approach reduces the number of kernel arguments while preserving the original data-parallel execution structure.

\subsection{Execution Mapping on MPPA}

The OpenCL execution model is mapped onto the Kalray MPPA Coolidge2 architecture using a host--device cluster configuration. The execution mapping on the MPPA architecture is illustrated in Fig.~\ref{fig:DH}.

One compute cluster runs the OpenCL host, while the others execute device kernels. The planner consists of trajectory generation, collision checking, cost evaluation, and best trajectory selection, where the first three stages operate independently for each candidate. Each candidate trajectory is assigned to a work-item and distributed across compute clusters. Global memory stores shared data, while local and private memory are used for per-task computation.

\vspace{-8pt}
\section{Evaluation}\label{sec:evaluation}
This section evaluates the practicality of the proposed framework by
examining execution behavior on Kalray MPPA Coolidge2 platform
and comparing manually implemented and automatically generated code.

\begin{table}[t]
\centering
\caption{Experimental setup for Simulink-generated code}
\label{tab:setup_simulink}
\begin{tabular}{|l|l|}
\hline
Category & Description \\
\hline
Target processor & Kalray MPPA Coolidge2 \\
Compute clusters & 5 clusters (1 I/O CC + 4 compute CCs) \\
Operating system & Linux (on-board) \\
OpenCL runtime & Kalray OpenCL runtime \\
Baseline & Simulink C++ (Embedded Coder) \\
Planner & Frenet-frame-based trajectory planner \\
\hline
\textbf{Number of candidates} & \textbf{4,000 trajectories} \\
Lateral offset $d$ & $[-5.0, 4.75]$ m, step 0.25 m (40 samples) \\
Planning time $T$ & $[4.0, 6.25]$ s, step 0.25 s (10 samples) \\
Velocity samples & 10 samples around reference speed \\
\hline
Iterations & 45 planning cycles \\
Measurement metric & End-to-end planning time per call \\
Reported values & Median and maximum observed end-to-end execution \\ & time over 45 planning cycles \\
\hline
\end{tabular}
\vspace{-10pt}
\end{table}

\vspace{-6pt}

\subsection{Experimental Setup}

Performance was evaluated on Kalray MPPA Coolidge2 platform.
The main experimental conditions are summarized in Table~\ref{tab:setup_simulink}.
The Frenet planner was configured to generate a fixed number of candidate trajectories for each planning cycle. The execution time of the trajectory planning process was measured and compared with a CPU-based serial implementation.
All measurements were obtained by repeated executions under identical conditions.

\begin{figure}[t]
  \centering
  \includegraphics[width=0.85\textwidth,keepaspectratio]{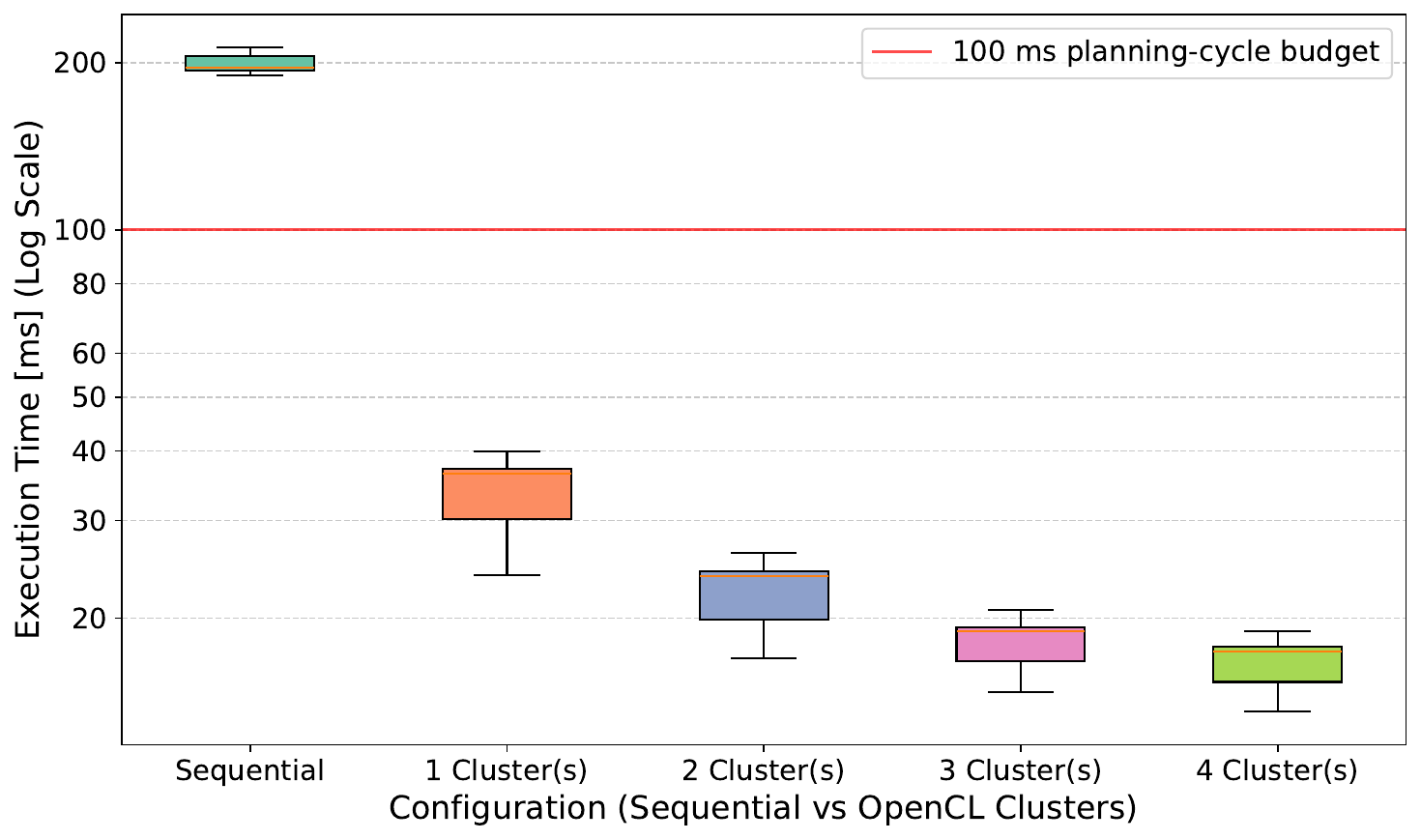}
  \caption{Execution time of Frenet-frame-based trajectory planning with 4,000 candidates on Kalray MPPA Coolidge2 under sequential execution and OpenCL-based cluster parallelization.}
  \label{fig:eval1}
  \vspace{-10pt}
\end{figure}

\vspace{-6pt}
\subsection{Performance Results}

The execution time of the trajectory planning task under the experimental conditions summarized in Table~\ref{tab:setup_simulink} is shown in Fig. ~\ref{fig:eval1}. The results compare sequential execution with OpenCL-based parallel execution using one host cluster and one to four compute clusters.
\vspace{-6pt}

\subsubsection{Absolute Performance Improvement}

In the sequential configuration generated by Embedded Coder, the median execution time was 195.1 ms, and the maximum observed end-to-end execution time reached 212.8 ms. \review{R.1.2}{This value exceeded the 100\,ms planning-cycle budget, which is derived from a typical LiDAR sensing period of 10\,Hz commonly adopted in autonomous driving systems~\cite{ros-lite,MBP2}, indicating that sequential model-based code did not meet the assumed timing target.}

When the proposed CUDA-to-OpenCL conversion workflow was applied and executed on four compute clusters, the median execution time was reduced to 17.4 ms, and the maximum observed end-to-end execution time was 18.9 ms. This result corresponds to approximately a 11.2× speedup compared with the sequential baseline executed on a single core.

The maximum observed end-to-end execution time under parallel execution remained below 20 ms, which is significantly lower than the assumed 100 ms planning-cycle budget. These results indicate that trajectory-level parallelism was effectively exploited on the many-core architecture while preserving the model-based workflow.

\vspace{-6pt}
\subsubsection{Cluster-Level Scalability}

Cluster-level scalability  was evaluated by increasing the number of compute clusters from one to four.

Execution time decreased from 36.4 ms with one cluster to 17.4 ms with four clusters, corresponding to a 2.1× speedup. The speedup was 1.5× for two clusters and 1.9× for three clusters relative to the single-cluster configuration, indicating diminishing returns as the number of clusters increases.

This behavior is attributed to fixed host-side overheads, such as kernel launch, synchronization, and command queue management, which become more dominant as the workload per cluster decreases.

\vspace{-6pt}

\subsection{Maximum Processable Candidates Under 100 ms Planning-Cycle Budget}

\begin{figure}[t]
  \centering
  \includegraphics[width=0.85\textwidth,keepaspectratio]{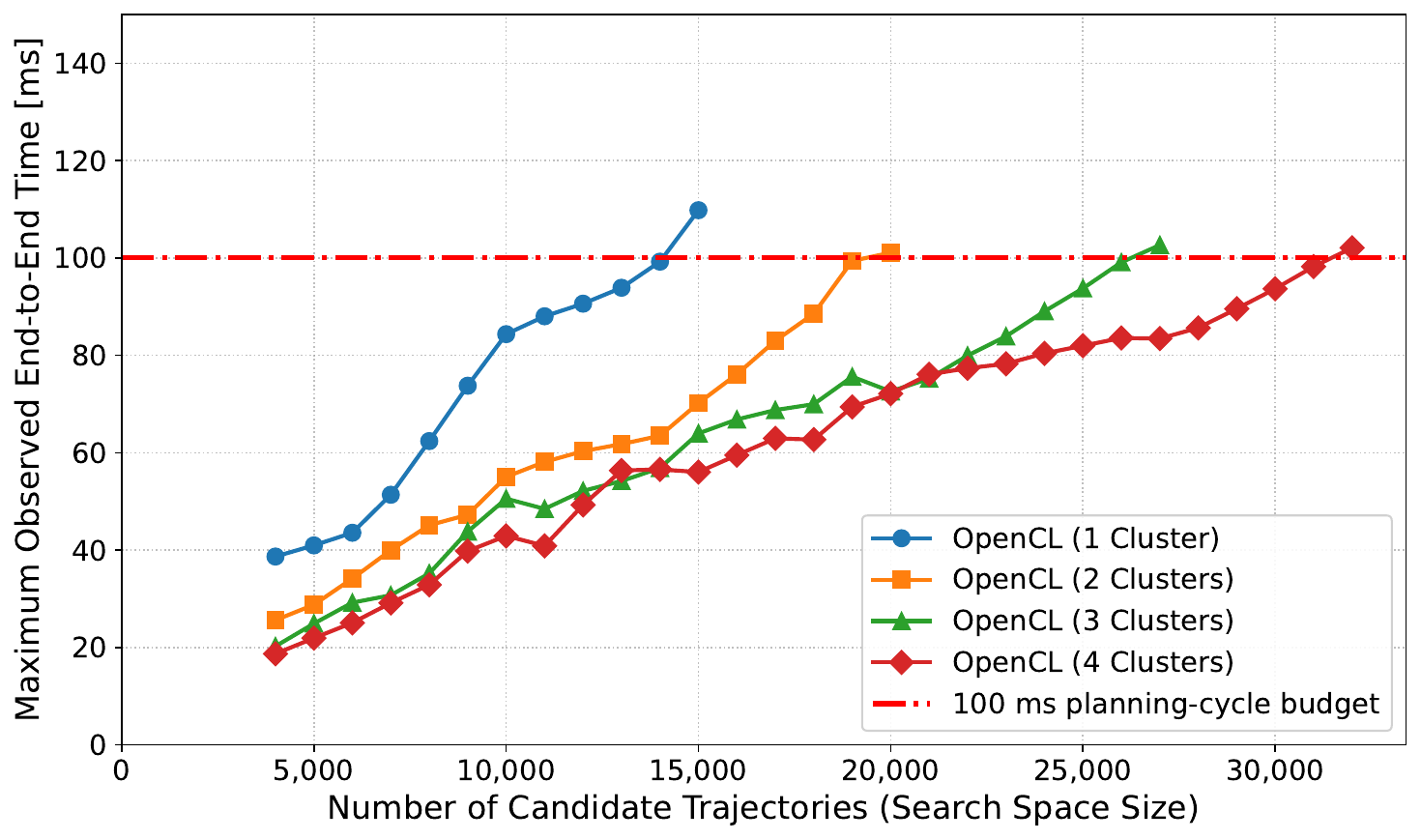}
  \caption{Maximum observed end-to-end execution time over 45 planning cycles versus the number of candidate trajectories for one to four compute clusters. The dashed line indicates the assumed 100 ms planning-cycle budget.}
  \label{fig:eval2}
      \vspace{-10pt}

\end{figure}

To evaluate the computational limit under the assumed timing condition, the number of candidate trajectories was increased while measuring the execution time for one to four compute clusters. The experimental conditions were identical to those in Table~\ref{tab:setup_simulink}, except that the lateral offset sampling range ($d$) was expanded to generate more candidates.

The number of trajectories was increased in increments of 1,000 by
adjusting the lateral offset sampling parameter $d$.
The reported thresholds correspond to the first trajectory count at which the maximum observed end-to-end execution time exceeded the assumed 100 ms planning-cycle budget.
Therefore, the maximum number of trajectories that remains within the budget is 1,000 less than the reported threshold.

The maximum number of trajectories that can be processed within
100 ms was 14,000 for one cluster, 19,000 for two clusters,
26,000 for three clusters, and 31,000 for four clusters.

Compared with the single-cluster configuration, the four-cluster
configuration increased the exploration capacity from 14,000 to
31,000 candidate trajectories, corresponding to a 2.21$\times$
increase while remaining within the assumed timing threshold.
These results indicate that the proposed trajectory-level parallelization increases the feasible search-space size under the evaluated planning-cycle budget and runtime configuration.

\subsection{Runtime Granularity and OS-Level Effects}

During early profiling experiments, execution times were quantized in approximately 10 ms increments. This phenomenon was traced to the Linux kernel timer configuration (CONFIG\_HZ) and the completion-wait strategy used in the OpenCL runtime, which relied on sleep-based polling.

Timer-driven completion detection masked fine-grained device-side scalability by rounding measured execution times to the next scheduling tick. As a result, scalability improvements beyond certain thresholds were not accurately reflected in the initial measurements.

To eliminate this runtime-induced quantization effect, the POCL runtime was reconfigured to use a busy-wait polling strategy for kernel completion detection. After this modification, the 10 ms granularity constraint was removed, enabling fine-grained measurement of execution behavior.

The refined measurements revealed consistent performance improvements as the cluster count increased. These observations confirm that the scalability limitation observed in early experiments originated from runtime scheduling behavior rather than hardware constraints. Accurate evaluation of many-core scalability therefore requires careful control of operating system scheduling policies and runtime configuration.

\vspace{-8pt}
\subsection{Lessons Learned}
This evaluation yields four main lessons.

\textbf{Lesson 1 (RQ1): Model structuring and workload characteristics jointly determine retargetability to many-core platforms.}
Generating parallel code from a Simulink model requires that computationally intensive logic be implemented in MATLAB Function blocks so that GPU Coder can emit CUDA kernels. Furthermore, the workload must be decomposable into independent candidate-wise or data-wise tasks. The Frenet planner satisfies these conditions because trajectory generation, collision checking, and cost evaluation are independent across candidates, enabling a direct one-work-item-per-candidate mapping. These conditions are not specific to the Frenet planner, but apply to a broader class of CPS workloads such as candidate-based trajectory evaluation, independent scenario evaluation, and data-parallel state or cost updates.

\textbf{Lesson 2 (RQ2): Workflow-preserving retargeting requires platform-aware translation beyond syntax rewriting.} While CUDA and OpenCL share similar execution models, preserving the generated host/device structure requires additional mechanisms.
Host-side orchestration, memory transfers, synchronization, and result collection can dominate total latency as device-side execution time decreases with more clusters. Scalability is therefore not perfectly linear; runtime and OS configuration (e.g., sleep-based completion detection) must be controlled to expose the actual scaling trend. For the evaluated 4,000-candidate workload, the end-to-end time remained within the 100\,ms budget on all cluster counts tested.

\textbf{Lesson 3: End-to-end latency, not kernel execution time alone, governs the observed timing behavior.}
 The total latency of a planning cycle includes host-side orchestration, memory transfers, kernel execution, synchronization, and result collection. In the evaluated setting, sequential execution exceeds the 100 ms planning-cycle budget, whereas the OpenCL implementation satisfies it with significant margin. Moreover, the maximum admissible number of candidate trajectories depends on the relationship between search-space size, platform configuration, and runtime overhead. Thus, the planning-cycle budget should be treated as a parameter that constrains the feasible workload size rather than a fixed target.

\textbf{Lesson 4: Scalability is bounded by runtime and system-level overheads rather than compute capacity alone.}
Increasing the number of compute clusters reduces execution time, but the speedup is not linear due to fixed overheads such as kernel launch, command queue management, synchronization, and memory transfer. In addition, runtime and operating system behaviors, such as timer granularity and polling strategies, can significantly affect observed performance. Accurate evaluation of many-core CPS execution therefore requires considering both architectural scalability and runtime-level effects.
\label{sec:lessons_learned}

\vspace{-8pt}
\section{Related Work}\label{sec:related}

\begin{table}[t]
\centering
\caption{Comparison with related methods and features}
\label{tab:related_work}
\scriptsize
\setlength{\tabcolsep}{3pt}
\renewcommand{\arraystretch}{1.2}

\begin{tabular}{lcccccccc}
\hline
 & MBD & ACG & CU & OC & MC & DP & TP & WP \\
\hline
MBP~\cite{MBP1,MBP2,MBP3}                  
& \checkmark & \checkmark &  &  & \checkmark &  & \checkmark &  \\

ROS 2 PCG~\cite{data-parallel}            
& \checkmark & \checkmark &  &  &  &  & \checkmark &  \\

Frenet GPU~\cite{MUZZINI2024103239}       
&  & \checkmark & \checkmark &  &  & \checkmark &  &  \\

CU2CL~\cite{martinez2011cu2cl}            
&  &  & \checkmark & \checkmark &  &  &  &  \\

Swan~\cite{harvey2011swan}                
&  &  & \checkmark & \checkmark &  &  &  &  \\

OpenCL FW~\cite{6113788}                  
&  &  &  & \checkmark & \checkmark &  &  &  \\

\hline
This study                                
& \checkmark & \checkmark & \checkmark$^*$ & \checkmark & \checkmark & \checkmark & \checkmark & \checkmark \\
\hline
\end{tabular}

\vspace{2mm}
\footnotesize{
MBD: Model-Based Development,
ACG: Automatic Code Generation,
CU: CUDA,
OC: OpenCL,
MC: Many-core,
DP: Data Parallel,
TP: Task Parallel,
WP: Workflow Preservation.
\\
$^*$ CUDA is used as an intermediate representation and translated to OpenCL.
}
\vspace{-10pt}
\end{table}

As summarized in Table~\ref{tab:related_work}, prior work on parallelization and code generation can be broadly categorized into CUDA-to-OpenCL translation, high-level intermediate representations, and model-based parallelization approaches.

\subsection{CUDA-to-OpenCL Translation}

A substantial body of work has addressed source-level translation from CUDA to OpenCL.
CU2CL~\cite{martinez2011cu2cl} performs Clang-based abstract syntax tree (AST) analysis to convert general-purpose CUDA programs into OpenCL.
Similarly, Swan~\cite{harvey2011swan} provides a lightweight Python-based converter for translating CUDA code into OpenCL.
Du et al.~\cite{du2012cuda} investigated performance portability between CUDA and OpenCL and showed that API-level translation can achieve comparable performance across platforms.

However, these approaches target general-purpose CUDA programs and do not consider code generated from model-based development tools such as Simulink.
Moreover, they do not address platform-specific constraints, such as kernel argument limits or cluster-level execution mapping on many-core processors.

\subsection{High-Level Intermediate Representations}

Recent approaches have explored higher-level abstractions to improve portability across heterogeneous platforms.
SYCL~\cite{kim2015sycl} provides a single-source C++ programming model that abstracts multiple backends, while AMD HIP~\cite{hip2024} offers a translation layer targeting GPU portability.
Although these approaches improve cross-platform support, they rely on specific runtime environments and are not directly applicable to platforms such as Kalray MPPA.

\subsection{Model-Based Parallelization}

Model-based approaches have been proposed to automatically generate parallel code from Simulink models.
Model-Based Parallelizer (MBP)~\cite{MBP1,MBP2,MBP3} extracts task-level parallelism from Simulink block structures for multicore execution.
In addition, recent work on parallelized code generation for ROS~2 nodes~\cite{data-parallel} enables event-driven and timer-driven execution by exploiting parallelism in Simulink-based models.

However, these approaches focus on parallelization within a target platform and do not address retargeting generated code across different parallel architectures or programming models.

\subsection{Summary of Differences}

Unlike prior CUDA-to-OpenCL translators such as CU2CL and Swan, the proposed framework targets GPU-Coder-generated CUDA and preserves the Simulink-based development workflow.
Furthermore, it addresses platform-specific constraints such as kernel argument limits and cluster-level execution mapping on many-core processors.

Compared with model-based parallelization approaches, this work focuses on workflow-preserving retargeting across heterogeneous many-core platforms rather than parallelization within a single platform.
\vspace{-8pt}
\section{Conclusion}\label{sec:conc}

This paper presented a workflow for retargeting GPU-Coder-generated CUDA code from a Simulink trajectory planning model to OpenCL for execution on a cluster-based many-core processor. The approach preserves the Simulink-based model-based development flow while enabling candidate-parallel execution of the trajectory planner through automatic translation of the supported CUDA subset into OpenCL host and kernel code. The generated implementation was deployed and evaluated on the Kalray MPPA Coolidge2 platform using a Frenet trajectory planning workload.

\textbf{RQ1} Answer: Retargetability requires expressing dominant computation in MATLAB Function blocks and decomposing it into independent candidate- or data-wise tasks, enabling GPU Coder to emit host/device code mappable to many-core execution. 
\textbf{RQ2} Answer: Workflow-preserving retargeting requires platform-aware mechanisms, including NDRange mapping, runtime API emulation, and argument packing, beyond syntax rewriting to preserve the generated execution structure under platform constraints. 
\textbf{RQ3} Answer: Feasibility is governed by end-to-end latency rather than kernel time alone, making admissible search-space size a function of the planning-cycle budget.
\textbf{RQ4} Answer: Latency decreases with cluster count but exhibits diminishing returns due to fixed overheads in host orchestration, synchronization, and OS/runtime behavior.

Overall, the study demonstrates that GPU-Coder-generated CUDA can serve as an intermediate representation for retargeting Simulink-based trajectory planners to OpenCL-based many-core processors. The results also highlight the importance of model structuring, workflow-level latency evaluation, and runtime configuration when deploying model-generated parallel code on cluster-based many-core architectures. 

\review{R.2.2}{Although the evaluation in this paper focuses on a Frenet trajectory planner, the proposed workflow is not inherently limited to this application. While the internal mechanisms of GPU Coder are proprietary, empirical observations of its output confirm that the generated CUDA code follows certain structural regularities, primarily relying on a limited subset of fixed patterns (e.g., standard kernel qualifiers, memory API calls, and launch syntax). The converter is designed to target these observed patterns rather than application-specific constructs. While our tool does not exhaustively guarantee coverage of all possible code permutations produced by GPU Coder, a Simulink model whose computationally intensive logic is expressed in MATLAB Function blocks and translated into this common CUDA subset is a candidate for the same Simulink-to-CUDA-to-OpenCL retargeting workflow. Validating this generalizability across additional application domains remains future work.}

\begin{credits}
\subsubsection{\ackname}
This work was supported by JST, FOREST Program, Grant Number JPMJFR242G.
\vspace{-8pt}
\end{credits}
\bibliographystyle{splncs04}
\bibliography{bstcontrol,Reference}

@INPROCEEDINGS{8443742,
  author={Kato, Shinpei and Tokunaga, Shota and Maruyama, Yuya and Maeda, Seiya and Hirabayashi, Manato and Kitsukawa, Yuki and Monrroy, Abraham and Ando, Tomohito and Fujii, Yusuke and Azumi, Takuya},
  booktitle={Proc. of {ACM}/{IEEE} International Conference on Cyber-Physical Systems (ICCPS)},
  title={Autoware on Board: Enabling Autonomous Vehicles with Embedded Systems}, 
  year={2018},
  volume={},
  number={},
  pages={287-296},
 }

@article{miura2021cosam,
  title={{CoSAM}: {Co-Simulation} Framework for {ROS}-based Self-driving Systems and {MATLAB}/{Simulink}},
  author={Miura, Keita and Tokunaga, Shota and Horita, Yuki and Oda, Yasuhiro and Azumi, Takuya},
  journal={Journal of Information Processing},
  volume={29},
  pages={227--235},
  year={2021},
  publisher={Information Processing Society of Japan}
}

@article{burgio2017software,
  title={A software stack for next-generation automotive systems on many-core heterogeneous platforms},
  author={Burgio, Paolo and Bertogna, Marko and Capodieci, Nicola and Cavicchioli, Roberto and Sojka, Michal and Houdek, P{\v{r}}emysl and Marongiu, Andrea and Gai, Paolo and Scordino, Claudio and Morelli, Bruno},
  journal={Microprocessors and Microsystems},
  volume={52},
  pages={299--311},
  year={2017},
  publisher={Elsevier}
}

@INPROCEEDINGS{werling2010,
  author={Werling, Moritz and Ziegler, Julius and Kammel, Sören and Thrun, Sebastian},
  booktitle={Proc. of  IEEE International Conference on Robotics and Automation}, 
  title={Optimal trajectory generation for dynamic street scenarios in a {Frenét Frame}}, 
  year={2010},
  volume={},
  number={},
  pages={987-993},
}

@INPROCEEDINGS{MBP1,
  author={Obi, Kenshin and Onozawa, Takumi and Fujimoto, Hiroshi and Azumi, Takuya},
  booktitle={Proc. of IEEE International Conference on Emerging Technologies and Factory Automation (ETFA)}, 
  title={Model-Based Development for Autonomous Driving Software Considering Parallelization}, 
  year={2024},
  volume={},
  number={},
  pages={1-8},
}

@ARTICLE{MBP2,
  author={Obi, Kenshin and Onozawa, Takumi and Yoshinaka, Ryo and Fujimoto, Hiroshi and Azumi, Takuya},
  journal={IEEE Open Journal of the Industrial Electronics Society}, 
  title={Model-Based Development for Event-Driven and Timer-Driven {ROS 2} Nodes Considering Parallelization}, 
  year={2026},
  volume={7},
  number={},
  pages={348-368},
}

@INPROCEEDINGS{MBP3,
  author={Zhong, Zhaoqian and Edahiro, Masato},
  booktitle={Proc. of International SoC Design Conference (ISOCC)}, 
  title={Model-based Parallelization for {Simulink} Models on Multicore {CPUs} and {GPUs}}, 
  year={2019},
  volume={},
  number={},
  pages={103-104},
}

@article{MUZZINI2024103239,
title = {{GPU} implementation of the {Frenet} Path Planner for embedded autonomous systems: A case study in the {F1tenth} scenario},
journal = {Journal of Systems Architecture},
volume = {154},
pages = {103239},
year = {2024},
issn = {1383-7621},
url = {https://www.sciencedirect.com/science/article/pii/S1383762124001760},
author = {Filippo Muzzini and Nicola Capodieci and Federico Ramanzin and Paolo Burgio},
}

@INPROCEEDINGS{6113788,
  author={Lee, Jun and Kim, Jungwon and Kim, Junghyun and Seo, Sangmin and Lee, Jaejin},
  booktitle={Proc. of International Conference on Parallel Architectures and Compilation Techniques}, 
  title={An {OpenCL} Framework for Homogeneous Manycores with No Hardware Cache Coherence}, 
  year={2011},
  volume={},
  number={},
  pages={56-67},
}

@article{werling2012optimal,
  title={Optimal trajectories for time-critical street scenarios using discretized terminal manifolds},
  author={Werling, Moritz and Kammel, S{\"o}ren and Ziegler, Julius and Gr{\"o}ll, Lutz},
  journal={The International Journal of Robotics Research},
  volume={31},
  number={3},
  pages={346--359},
  year={2012},
  publisher={SAGE Publications Sage UK: London, England}
}

@ARTICLE{broy2007,
  author={Broy, Manfred and Kruger, Ingolf H. and Pretschner, Alexander and Salzmann, Christian},
  journal={Proceedings of IEEE}, 
  title={Engineering Automotive Software}, 
  year={2007},
  volume={95},
  number={2},
  pages={356-373}
}

@article{nozaki2025dedicated,
  title={Dedicated Processor Allocation Scheduling for High-load Tasks on Clustered Many-core Processors},
  author={Nozaki, Yutaro and Okamura, Ryo and Koike, Ryotaro and Azumi, Takuya},
  journal={Journal of Information Processing},
  volume={33},
  pages={79--90},
  year={2025},
  publisher={Information Processing Society of Japan}
}

@INPROCEEDINGS{yabe,
  author={Yabe, Takuma and Azumi, Takuya},
  booktitle={Proc. of ACM/IEEE International Conference on Cyber-Physical Systems (ICCPS)}, 
  title={Exploring the Performance of Deep Neural Networks on Embedded Many-Core Processors}, 
  year={2022},
  volume={},
  number={},
  pages={193-202},
}

@inproceedings{martinez2011cu2cl,
  title={{CU2CL}: A {CUDA}-to-{OpenCL} Translator for Multi-and Many-Core Architectures},
  author={Martinez, Gabriel and Gardner, Mark and Feng, Wu-chun},
  booktitle={Proc. of IEEE International Conference on Parallel and Distributed Systems (ICPADS)},
  pages={300--307},
  year={2011},
}

@article{du2012cuda,
  title={From {CUDA} to {OpenCL}: Towards a Performance-portable Solution for Multi-platform {GPU} Programming},
  author={Du, Peng and Weber, Rick and Luszczek, Piotr and Tomov, Stanimire and Peterson, Gregory and Dongarra, Jack},
  journal={Parallel Computing},
  volume={38},
  number={8},
  pages={391--407},
  year={2012},
  publisher={Elsevier}
}

@article{harvey2011swan,
  title={{Swan}: A tool for porting {CUDA} programs to {OpenCL}},
  author={Harvey, Michael J and De Fabritiis, Gianni},
  journal={Computer Physics Communications},
  volume={182},
  number={4},
  pages={1093--1099},
  year={2011},
  publisher={Elsevier}
}

@inproceedings{kim2015sycl,
  title={{SYCL}-bench: A Versatile Cross-platform Benchmark Suite for Heterogeneous Computing},
  author={Lal, Sohan and Alpay, Aksel and Salzmann, Philip and Kreutzer, Moritz and Bode, Thomas and Hager, Georg and Wellein, Gerhard and Heuveline, Vincent},
 booktitle={Proc. of International European Conference on Parallel and Distributed Computing (Euro-Par)} ,
 pages={629--644},
  year={2023},
  publisher={Springer}
}

@misc{hip2024,
  author={{AMD}},
  title={{HIP}: Heterogeneous-Computing Interface for Portability},
  howpublished={\url{https://github.com/ROCm/HIP}},
  year={2024},
  note={Accessed 2026-04}
}

@INPROCEEDINGS{ros-lite,
  author={Azumi, Takuya and Maruyama, Yuya and Kato, Shinpei},
  booktitle={Proc. of IEEE/RSJ International Conference on Intelligent Robots and Systems (IROS)}, 
  title={{ROS}-lite: {ROS} Framework for {NoC}-Based Embedded Many-Core Platform}, 
  year={2020},
  volume={},
  number={},
  pages={4375-4382},
}

@article{derler2012cyber,
  title={Cyber-physical systems: Design challenges},
  author={Derler, Patricia and Lee, Edward A and Sangiovanni Vincentelli, Alberto},
  journal={Proc. of the IEEE},
  volume={100},
  number={1},
  pages={13--28},
  year={2011},
  publisher={IEEE}
}

@book{alur2015principles,
  title={Principles of cyber-physical systems},
  author={Alur, Rajeev},
  year={2015},
  publisher={MIT press}
}

@INPROCEEDINGS{data-parallel,
  author={Obi, Kenshin and Yoshinaka, Ryo and Fujimoto, Hiroshi and Azumi, Takuya},
  booktitle={Proc. of Euromicro Conference on Software Engineering and Advanced Applications (SEAA)}, 
  title={Parallelized Code Generation from {Simulink} Models for Event-driven and Timer-driven {ROS} 2 Nodes}, 
  year={2024},
  volume={},
  number={},
  pages={48-55},
}

\end{document}